\documentclass[showpacs,preprintnumbers,amsmath,amssymb,superscriptaddress]{revtex4}
\usepackage{amssymb}

\usepackage{amsmath,amssymb,graphicx,bm}

\begin{document}

\title{New types of $f(T)$ gravity}

\author{Rong-Jia Yang}
\email{yangrj08@gmail.com}
 \affiliation{College of Physical Science and Technology,
Hebei University, Baoding 071002, China}

\begin{abstract}
Recently $f(T)$ theories based on modifications of teleparallel gravity where torsion is the geometric object describing gravity instead of curvature have been proposed to explain the present cosmic accelerating expansion. The field equations are always second order, remarkably simpler than $f(R)$ theories. In analogy to the $f(R)$ theory, we consider here three types of $f(T)$ gravity, and find that all of them can give rise to cosmic acceleration with interesting features, respectively.
\end{abstract}

\makeatletter
\def\@pacs@name{PACS: 04.50.Kd, 95.36.+x, 98.80.-k}
\makeatother

\maketitle

\section{Introduction}
In the last decade a convergence of independent cosmological
observations suggested that the Universe is experiencing accelerated
expansion. Usually, an unknown energy component, dubbed dark energy, is
proposed to explain this acceleration. Dark energy almost evenly
distributes in the Universe, and its pressure is negative. The simplest and most theoretically appealing candidate of
dark energy is the vacuum energy (or the cosmological constant
$\Lambda$) with a constant equation of state (EoS) parameter $w=-1$.
This scenario is in general agreement with the current astronomical
observations, but has difficulties to reconcile the small
observational value of dark energy density with estimates from
quantum field theories; this is the cosmological constant problem \cite{weinberg}.
Recently it was shown that $\Lambda$CDM model may also suffer from an age problem \cite{Yang2010}.
It is thus natural to pursue alternative possibilities to explain
the mystery of dark energy. Over the past decade numerous dark energy models have been proposed,
such as quintessence, phantom, k-essence, tachyon, (generalized) Chaplygin gas, DGP, etc.
Rather than introduce a dark energy, one could propose modifications of the Einstein-Hilbert Lagrangian by adopting
different functions of the Ricci scalar, known as $f(R)$ theories (for reviews see e. g. \cite{Capozziello1,Felice, Sotiriou10, Capozziello, Silvestri, Nojiri07}), as a way to obtain a late accelerated expansion. Studies of the physics of $f(R)$ theories are however hampered by the complexity of the fourth order field equations in the framework of metric. Although the Palatini variational approach for such $f(R)$ theories leads to second order field equations, one still has difficulty to obtain both exact and numerical solutions which can be compared with observations in many cases. Recently, models based on modified teleparallel gravity were proposed as an alternative to $f(R)$ theories \cite{Bengochea}, namely $f(T)$ theories, in which the torsion will be responsible for the late accelerated expansion, and the field equations will always be second order equations. Originally $f(T)$ theories had been proposed as models for inflation \cite{Ferraro,Ferraro2007}.

Recently, $f(T)$ theories have been studied in extenso. Some $f(T)$ theories had been proposed in \cite{Bengochea,Linder, Myrzakulov,Bamba}. It had been shown that $f(T)$ theories are not dynamically equivalent to teleparallel action plus a scalar field via conformal transformation \cite{yang}. Observational constraints had been considered in \cite{Bengochea1,wu,Zhang,Wei}. Large-scale structure in $f(T)$ gravity had been discussed in \cite{Li2011}. Cosmological perturbations in $f(T)$ gravity had been investigated in \cite{Dent,Chen,Zheng,Cai}. In \cite{Meng}, Birkhoff's theorem in $f(T)$ gravity had been studied. Static solutions with spherical symmetry in $f(T)$ theories have been discussed in \cite {Wang}. Relativistic Stars in f(T) gravity had been investigated in \cite{Deliduman}. In \cite{Capozziello2011}, the cosmic expansion was studied by using cosmography. Although $f(T)$ gravity has attracted so much attentions, it had been pointed out that the action and the field equations of $f(T)$ are not invariant under local Lorentz transformations \cite{Li}, and it had been shown why restoring local Lorentz symmetry in such theories cannot lead to sensible dynamics, even if one gives up teleparallelism \cite{Sotiriou}. Due to the lack of local Lorentz invariance, $f(T)$ theories appear to have extra degrees of freedom with respect to general relativity \cite{Li,Sotiriou}, and the autoparallel frames satisfying the field equations are evasive to an a priori physical understanding \cite{Ferraro2011}. Though Lorentz symmetry has been experimentally tested, its violation is still possible \cite{Saveliev}, while teleparallel gravity and $f(T)$ theories afford possible choices to construct Lorentz violation theories. So $f(T)$ theories are worth further depth studies.

In this paper we will investigate three new types of $f(T)$ gravity, study their cosmological behaviors by using numerical methods, and discuss their potential physical implications. The paper is organized as follows, in the following section, we review $f(T)$ theories. In Sec. III, we study the proposed $f(T)$ gravities. Finally, we shall close with a few concluding remarks in Sec. IV.

\section{$f(T)$ gravities}
$f(T)$ theories based on modified teleparallel gravity. The teleparallel approach by using the vierbein as dynamical object was taken by Einstein \cite{Einstein,Einstein1930}. In teleparallel gravity, rather than use the curvature defined via the Levi-Civita connection, one could explore the model of torsion via the Weitzenb\"{o}ck connection, which has no curvature to describe gravity,
\begin{align}
\label{torsion2}
T^\lambda_{\:\:\mu\nu}\equiv e^\lambda_i(\partial_\mu e^i_\nu-\partial_\nu e^i_\mu),
\end{align}
where $e_i^\mu$ ($\mu=0, 1, 2, 3$) are the components of the vierbein field ${\mathbf{e}_i(x^\mu)}$ ($i=0, 1, 2, 3$) in a coordinate basis, i.e. $\mathbf{e}_i=e^\mu_i\partial_\mu$.
The vierbein is an orthonormal basis
for the tangent space at each point $x^\mu$ of the manifold: $\mathbf{e}%
_i\cdot\mathbf{e}_j=\eta_{i\, j}$, where $\eta_{i\, j}=$diag $(1,-1,-1,-1)$. Notice that Latin indices refer to the tangent space, while Greek indices
label coordinates on the manifold. The metric tensor is obtained from the
dual vierbein as $g_{\mu\nu}(x)=\eta_{i\, j}\, e^i_\mu (x)\, e^j_\nu (x)$.

The teleparallel Lagrangian is \cite{Hayashi, Maluf, Arcos}
\begin{equation}  \label{lagTele}
T\equiv S_\rho^{\:\:\:\mu\nu}\:T^\rho_{\:\:\:\mu\nu},
\end{equation}
where
\begin{equation}  \label{S}
S_\rho^{\:\:\:\mu\nu}=\frac{1}{2}\Big(K^{\mu\nu}_{\:\:\:\:\rho}+\delta^\mu_%
\rho \:T^{\theta\nu}_{\:\:\:\:\theta}-\delta^\nu_\rho\:
T^{\theta\mu}_{\:\:\:\:\theta}\Big),
\end{equation}
and the contorsion tensor, $K^{\mu\nu}_{\:\:\:\:\rho}$, is
\begin{equation}  \label{K}
K^{\mu\nu}_{\:\:\:\:\rho}=-\frac{1}{2}\Big(T^{\mu\nu}_{\:\:\:\:\rho}
-T^{\nu\mu}_{\:\:\:\:\rho}-T_{\rho}^{\:\:\:\:\mu\nu}\Big),
\end{equation}
which equals the difference between Weitzenb\"{o}ck and Levi-Civita
connections.

Following Ref. \cite{Bengochea} we promote the teleparallel Lagrangian density as a function of $T$, in analogy to $f(R)$ theories.
Thus the action reads
\begin{equation}  \label{accionTP}
I= \frac{1}{16\, \pi\, G}\, \int d^4x\:e\:f(T)+I_{\rm m},
\end{equation}
where $e=det(e^i_\mu)=\sqrt{-g}$. The variation of the action with respect to the vierbein leads to the field equations
\begin{eqnarray}
[e^{-1}\partial_\mu(e\:S_i^{\:\:\:\mu\nu})-e_i^{\:\lambda}
\:T^\rho_{\:\:\:\mu\lambda}\:S_\rho^{\:\:\:\nu\mu}]f_T+ S_i^{\:\:\:\mu\nu}\partial_\mu T f_{TT}
+\frac{1}{4}
\:e_i^\nu \:f(T)=\frac{1}{2}k^2\:e_i^{\:\:\:\rho}\:T_\rho^{\:\:\:\nu},  \label{ecsmovim}
\end{eqnarray}
where $k^2=8\pi G$, $f_T\equiv df/dT$, $f_{TT} \equiv d^2f/dT^2$, $S_i^{\:\:\mu\nu}\equiv e_i^{\:\:\rho}S_\rho^{\:\:\mu\nu}$, and
$T_{\mu\nu}$
is the matter energy-momentum tensor. Equations (\ref{ecsmovim}) are 2nd order, simpler than the dynamical equations resulting in
$f(R)$ theories.

For a flat homogeneous and isotropic FRW universe, one has
\begin{equation}  \label{tetradasFRW}
e^i_\mu=diag(1,a(t),a(t),a(t)),
\end{equation}
where $a(t)$ is the cosmological scale factor. By combining with (\ref{torsion2}), (\ref{S}) and (\ref{K}) one obtains
\begin{equation}  \label{STFRW}
T\equiv S^{\rho\mu\nu}T_{\rho\mu\nu}=-6\:H^2,
\end{equation}
where $H$ is the Hubble parameter $H=\dot{a}/a$. The modified Friedmann equations read
\begin{eqnarray}\label{f1}
	12H^2 f_{T}+f &=& 2k^2 \rho,\\
\label{f2}
	48 H^2 \dot{H}f_{TT}-(12H^2+4\dot{H})f_{T}-f &=& 2k^2p,
\end{eqnarray}
where $\rho$ and $p$ are the total density and pressure respectively. It must be point out that in the absence of Lorentz invariance there are
extra propagating degrees of freedom in $f(T)$ theories, thus, a FRW space-time (flat, open, or close) does not uniquely lead to the vierbein choice (\ref{tetradasFRW}) (for details see \cite{Li, Sotiriou,Ferraro2011}). Recently, Ref. \cite{Ferraro2011} argued that Eqs. (\ref{f1}) and (\ref{f2}) become a set of consistent dynamical equations for the vierbein choice (\ref{tetradasFRW}). If we require that the modified Friedmann equations (\ref{f1}) and (\ref{f2}) reduce to the Friedmann equations in general relativity when $f(T)$ theories reduce to general relativity ($f(T)=T$), we only have one unique vierbein choice: (\ref{tetradasFRW}). In this sense, the modified Friedmann equations (\ref{f1}) and (\ref{f2}) are the true equations.

The conservation equation is
\begin{eqnarray}
\label{rho}
\dot{\rho}+3H(\rho+p) &=& 0.
\end{eqnarray}
The modified Friedmann equations, (\ref{f1}) and (\ref{f2}), can be rewritten as
\begin{eqnarray}
\frac{3}{k^2}H^2 &=& \rho+\rho_T, \\
\frac{1}{k^2}(2\dot{H}+3H^2) &=& -(p+p_T),
\end{eqnarray}
where
\begin{eqnarray}
\label{rhoT}
\rho_T &=& \frac{1}{2k^2}(-12H^2f_T-f+6H^2), \\
\label{pT}
p_T &=& -\frac{1}{2k^2}[48\dot{H}H^2f_{TT}-4\dot{H}f_{T}+4\dot{H}]-\rho_T,
\end{eqnarray}
are the torsion contributions to the energy density and pressure. Then, by using Eqs. (\ref{rhoT}) and (\ref{pT}),
we can define the total and the effective torsion equations of state as
\begin{eqnarray}
w_{\rm tot} & \equiv & \frac{p+p_T}{\rho+\rho_T}=-1+\frac{2(1+z)}{3H}\frac{dH}{dz},\\
w_{\rm eff} & \equiv & \frac{p_T}{\rho_T}=-1-\frac{48\dot{H}H^2f_{TT}-4\dot{H}f_{T}+4\dot{H}}{-12H^2f_T-f+6H^2},
\end{eqnarray}
The deceleration parameter is defined as usual
\begin{eqnarray}
q \equiv -\frac{\ddot{a}}{aH^2}=-1+\frac{(1+z)}{H}\frac{dH}{dz}.
\end{eqnarray}
With the total equation of state or the deceleration parameter, we can determine whether there exists acceleration phase or not.

\section{New types of $f(T)$ gravity}
In this section, we will propose three types of $f(T)$ theory and discuss briefly their behavior in FRW cosmology. Thought, the forms of $f(T)$ theories considered here are similar with those of $f(R)$ theories, because of $R=6(2H^2+\dot{H})$ while $T=-6H^2$, the dynamic evolutions of $f(T)$ theories are different from those of the analogical $f(R)$ theories. In other words, $f(T)$ theories present new and interesting effects as compared to $f(R)$ theories. Other subjects, such as observational constraints and stability, will be considered elsewhere.

\subsection{Type I: $f(T)=T-\frac{\alpha}{(-T/6)^n}+\beta (-\frac{T}{6})^m$}
It has been shown that $f(R)=R-1/R$ model which is equivalent to some scalar-tensor gravity is ruled out as realistic theory due to the constraints to such Brans-Dicke type theories \cite{Chiba}. The cases, $f(R)=R-\alpha/R^n$, also have shown difficulties with weak field tests \cite{Olmo}, gravitational instabilities \cite{Dolgov}, and do not present a matter dominated era previous to the acceleration era \cite {Amendola}. However, by adding the scalar curvature squared term to the action, namely $f(R)=R-\alpha/R^n+\beta R^m$ \cite{Nojiri}, the mass of the scalar field can be adjusted to be very large and the scalar field can be decouple. Hence, this modified gravity theory passes the Solar System tests. In addition, this modified gravity has a newtonian limit which does not deviate significantly from the one in General Relativity. In analogy to this type $f(R)$ theory, we consider a type $f(T)$ theory: $f(T)=T-\frac{\alpha}{(-T/6)^n}+\beta (-\frac{T}{6})^m$. The motivations for this type of $f(T)$ theory are: firstly it is the simplest linear combination of two types of $f(T)$ theories, $\frac{\alpha}{(-T/6)^n}$ \cite{Bengochea} and $\beta (-\frac{T}{6})^m$ \cite{Linder}, so new dynamical effects can be expected; secondly when $T\longrightarrow \infty$, general relativity is recovered for $m\leq 1$ in early times (if $m> 1$, however, this type of $f(T)$ theory differs from general relativity even for $T\longrightarrow \infty$), while for later times the term $-\frac{\alpha}{(-T/6)^n}$ dominates and possibly leads to accelerated expansion. This type of $f(T)$ theory has an infinite cosmological constant in Minkowski space ($T=0$) unless $n=0$ (in which case it still has an explicit finite cosmological constant. Models with explicit cosmological constant, however, are generally not regarded as interesting since all the observations can be explained with just that one element, without the need to add terms coming from $f(T)$).

The modified Friedmann equation, (\ref{f1}), for this type $f(T)$ theory reads
\begin{eqnarray}
\label{y1}
y^n \left[y-\frac{1}{6}(2m-1)\beta H^{2(m-1)}_0 y^m-B \right]=C,
\end{eqnarray}
where $y\equiv H^2/H^2_0$, $B=\Omega_{\rm m0}(1+z)^3+\Omega_{\rm r0}(1+z)^4$, and $C=(2n+1)\alpha H^{-2(n+1)}_0/6$. This equation for $z=0$ allows us to rephrase the constant $C$ as $C=1-\Omega_{\rm m0}-\Omega_{\rm r0}-\frac{1}{6}(2m-1)\beta H^{2(m-1)}_0$. So the values of $\alpha$ and $\beta$ have been related together through Eq. (\ref{y1}). For $\alpha=0$, we obtain the model discussed in \cite{Linder}. For $\beta=0$, we obtain the model discussed in \cite{Bengochea}. The case $\beta=0$ and $n=0$ recovers the general relativity with cosmological constant, $\Omega_\Lambda=1-\Omega_{\rm m0}-\Omega_{\rm r0}$. An interesting point to be highlighted is that for $\alpha\neq 0$ and $\beta\neq 0$ the term $\beta (-T/6)^m$ contributes nothing to the dynamic behaves if $m=1/2$, just like the case $\alpha\neq 0$ and $\beta= 0$. Another interesting characteristic of this type $f(T)$ theory is that it can behave like a phantom, which affords an alternative realization to obtain $w_{\rm eff}<-1$.
These interesting characteristics are worthy of further detailed studies.

The effective energy density and pressure of torsion are given by
\begin{eqnarray}
\rho_T &=&\frac{1}{2k^2} \left[\frac{(2n+1)\alpha}{H_0^{2n}y^n} + (2m-1)\beta H_0^{2m}y^m \right],
\end{eqnarray}
\begin{eqnarray}
p_T = -\rho_T -\frac{\dot{H}}{3k^2} \times \left[\frac{-n(2n+1)\alpha}{H_0^{2(n+1)}y^{n+1}} + m(2m-1)\beta H_0^{2(m-1)}y^{m-1}\right].
\end{eqnarray}

Now we numerically investigate the cosmological behaves of this type of $f(T)$ theory. Figure 1 shows the evolution of the effective torsion equation of state $w_{\rm eff}$ as a function of $z$ for this type $f(T)$ theories with $\Omega_{\rm m0}=0.3$, $\Omega_{\rm r0}=8\times 10^{-5}$, $H_0=72$ km s$^{-1}$Mpc$^{-1}$, $m=0.2$, $n=0.1$, and $\beta H_0^{2(m-1)}=0.14$. $w_{\rm eff}$ run close $-1.1$ at high redshift and runs close $-1$ in the future. In this case, the model behaves like phantom. In the model with $f(R)=R-\alpha/R^n+\beta R^m$ \cite{Nojiri} the term $-\alpha/R^n$ dominates in later time, with a effective equation of state: $w_{\rm eff}\simeq -1+\frac{2(n+2)}{3(2n+1)(n+1)}$ \cite{Carroll2004}. According to their $w_{\rm eff}$, the difference between this type, the $f(T)$ theory and the analogical $f(R)$ theory is obvious. The evolution of the total equation of state is shown in Fig. 2. The latest three phases of the evolution of the universe: late acceleration ($w\rightarrow -1$), matter dominated ($w=0$), and radiation dominated ($w=1/3$), can be observed. Figure 3 shows the evolution of the deceleration parameter. The transition from deceleration to acceleration occurs at $z\sim 0.8$ in good agreement with recent observations \cite{Melchiorri}.

\begin{figure}
\includegraphics[width=10cm]{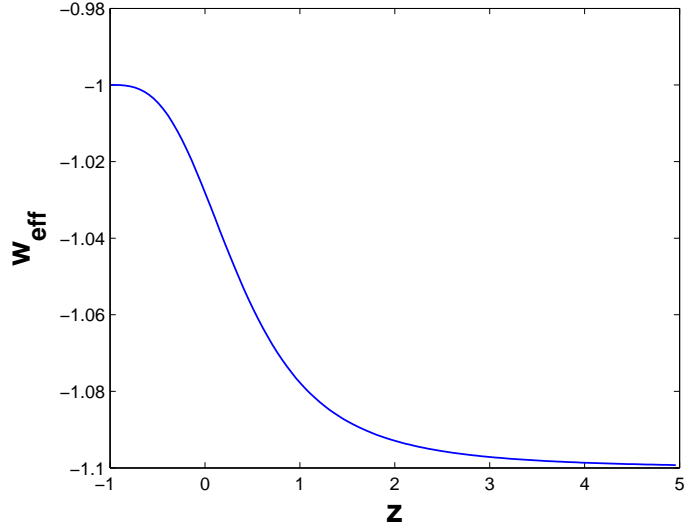}
\caption{The curve correspond to the effective torsion equation of state as a function of $z$ expected for type I $f(T)$ theories with $\Omega_{\rm m0}=0.3$, $\Omega_{\rm r0}=8\times 10^{-5}$, $H_0=72$ km s$^{-1}$Mpc$^{-1}$, $m=0.2$, $n=0.1$, and $\beta H_0^{2(m-1)}=0.14$. \label{Fig1}}
\end{figure}
\begin{figure}
\includegraphics[width=10cm]{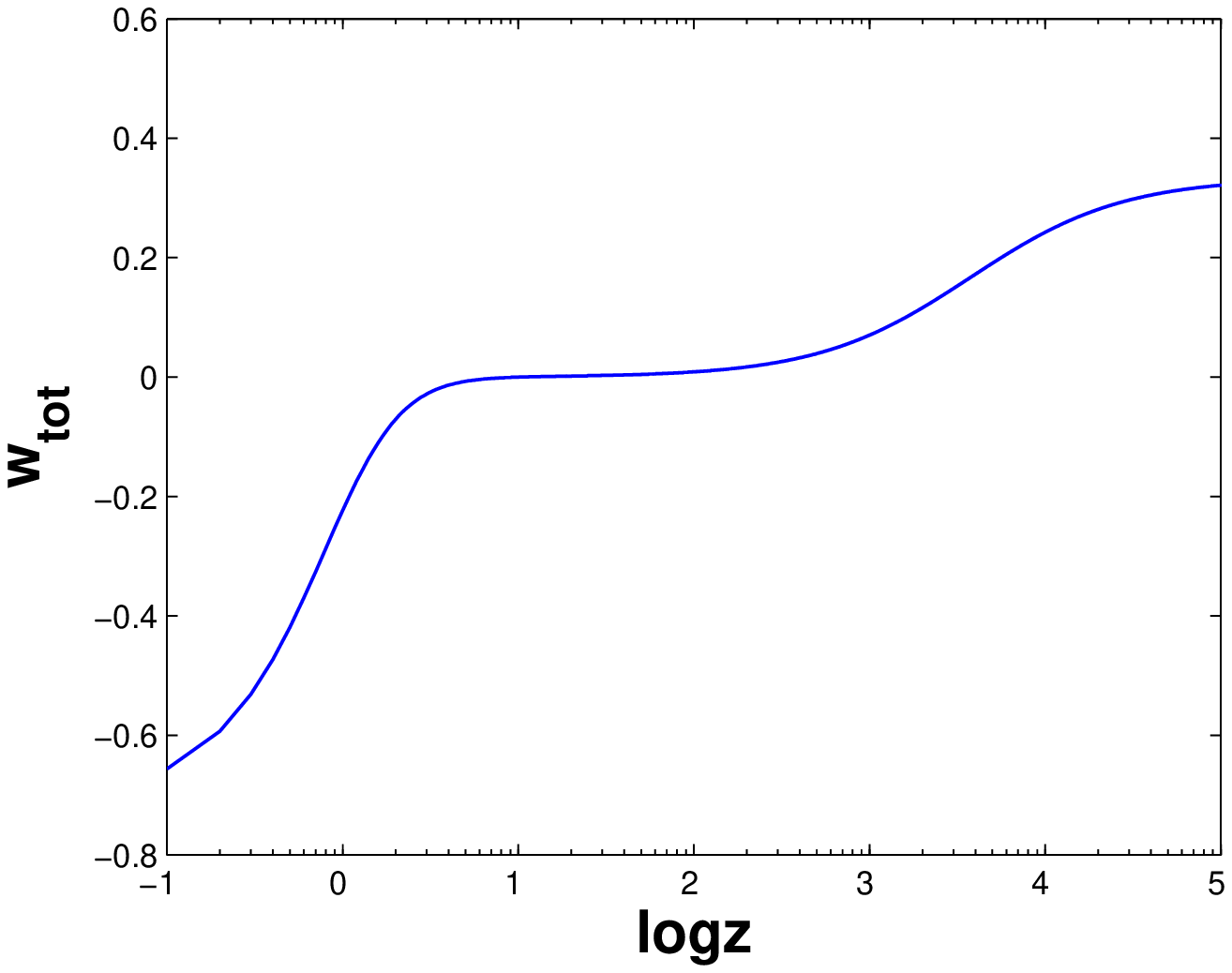}
\caption{The curve correspond to the total equation of state as a function of $z$ expected for type I $f(T)$ theories with the same values of the parameters in Fig. 1. \label{Fig2}}
\end{figure}
\begin{figure}
\includegraphics[width=10cm]{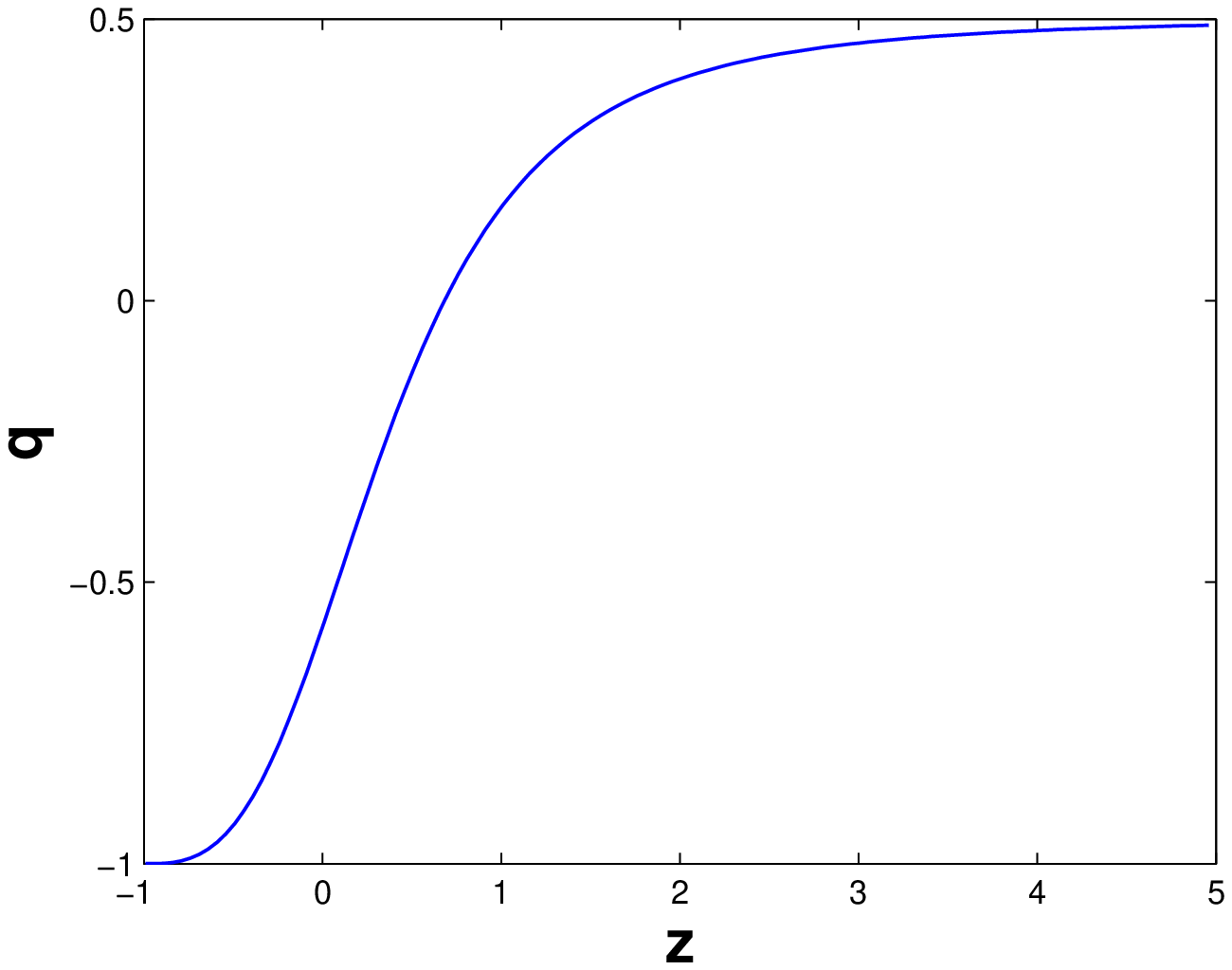}
\caption{The evolution of the deceleration parameter as a function of $z$ expected for type I $f(T)$ theories with the same values of the parameters in Fig. 1. \label{Fig3}}
\end{figure}

\subsection{Type II: $f(T)=T-\alpha T_0 \big[\big(1+\frac{T^2}{T^2_0} \big)^{-n}-1 \big]$}
 A $f(R)$ theory, which passed the observational and theoretical constraints, was proposed by Starobinsky \cite{Starobinsky}: $f(R)=R+\alpha R_0 \big[\big(1+\frac{R^2}{R^2_0} \big)^{-n}-1 \big]$, where $\alpha$, $R_0$, and $n$ are positive constants, $R_0$ is of the order of the present Ricci scalar. In analogy to this $f(R)$ theory, we consider another $f(T)$ theory: $f(T)=T-\alpha T_0 \big[\big(1+\frac{T^2}{T^2_0} \big)^{-n}-1 \big]$, with $\alpha$ and $n$ are positive constants, $T_0$ is the order of the present Hubble parameter. The motivation for this type of $f(T)$ theory like the case in the analogical $f(R)$ theory: firstly, for $f(0)=0$, the cosmological constant disappears in a flat space-time, thus the origin of dark energy can be regarded as the geometrical one; secondly, in the past or in the future, this type of $f(T)$ theory can behave like the general relativity with cosmological constant (but not a true cosmological constant). This interesting characteristic makes it possible to pass observational tests, while impossible to be distinguished from a $\Lambda$CDM model. It is well-known that $\Lambda$CDM suffers from coincidence problem. This type of $f(T)$ theory can behave very much like a cosmological constant, but because of its dynamic behavior, it is free from coincidence problem.  This is an interesting characteristic of this type of theory, worthy of further in-depth study.

The modified Friedmann equation, (\ref{f1}), for this type $f(T)$ theory reads
\begin{eqnarray}
\label{y2}
y+\frac{4n\alpha y^2}{(1+y^2)^{n+1}}+\frac{\alpha}{(1+y^2)^{n}}  -B=\alpha,
\end{eqnarray}
here we take $T_0=-6H^2_0$. This equation for $z=0$ allow us to rephrase the constant $\alpha$ as $\alpha=(1-\Omega_{\rm m0}-\Omega_{\rm r0})/(1-2^{-n+1} n-2^{-n})$. Compared with general relativity, $n$ is the sole new free parameter, since the specifying the value of $n$ and $\Omega_{\rm m0}$ ($\Omega_{\rm r0}$) the value of $\alpha$ is automatically fixed through Eq. (\ref{y2}). For $\alpha=0$ or $n=0$ one recovers the general relativity. The case $n\gg 1$ gives the general relativity with cosmological constant. Both at low redshift ($T\longrightarrow T_0$) and at high redshift ($T\longrightarrow \infty$), general relativity with cosmological constant is recovered. It should be pointed out that for $n=-1$ this type of $f(T)$ theory is a special case of type I $f(T)$ theory and is strongly different from general relativity.

The effective energy density and pressure of torsion are given by
\begin{eqnarray}
\rho_T &=&\frac{H^2_0}{2k^2} \left[\frac{-24n \alpha y^2}{(1+y^2)^{n+1}}- \frac{6\alpha}{(1+y^2)^{n}} + 6\alpha \right], \\
\label{p2}
p_T &=& \frac{8y\dot{H}}{2k^2} \left[\frac{4n(n+1)\alpha y^2}{(1+y^2)^{n+2}} - \frac{3n\alpha}{(1+y^2)^{n+1}} \right]-\rho_T.
\end{eqnarray}

Figure 4 shows the evolution of the effective torsion equation of state $w_{\rm eff}$ as a function of $z$ for this type $f(T)$ theories with $\Omega_{\rm m0}=0.3$, $\Omega_{\rm r0}=8\times 10^{-5}$, and $n=3.5$. The model acts like a cosmological constant at high redshift
and in the future; it only deviates slightly at present. For comparison, the effective equation of state $w_{\rm eff}$ of $f(R)=R+\alpha R_0 \big[\big(1+\frac{R^2}{R^2_0} \big)^{-n}-1 \big]$ is also shown in Fig \ref{StarobinskyF}. It is obvious that $w_{\rm eff}$ in the $f(R)$ model \cite{Starobinsky} can cross the phantom divide, while that of the analogical $f(T)$ theory cannot. Figure 6 shows the evolution of the total equation of state. The latest three phases of the evolution of the universe, late acceleration ($w\rightarrow -1$), matter dominated ($w=0$), and radiation dominated ($w=1/3$), can also be observed.
Figure 7 shows the evolution of the deceleration parameter.The transition from deceleration to acceleration occurs at $z\sim 1.2$, a little higher than the
transitional redshift constrained form observations, but one can adjust the parameters $\Omega_{\rm m0}$ and $n$ to obtain a lower transitional redshift. So it
is necessary to constrain this type of $f(T)$ theory with observations in future studies.

\begin{figure}
\includegraphics[width=10cm]{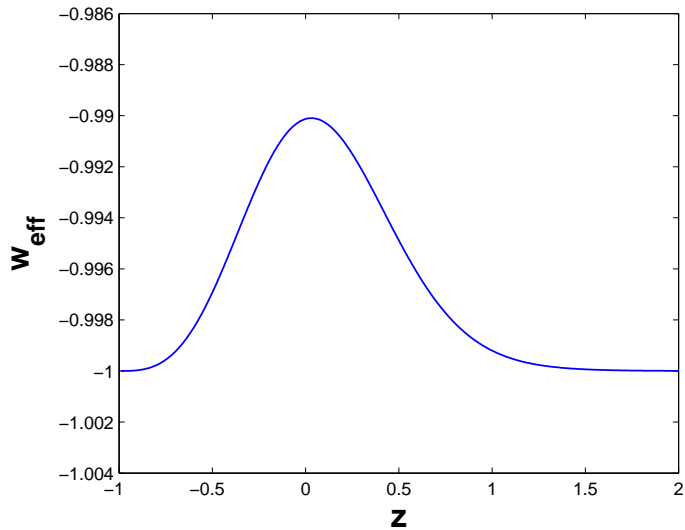}
\caption{The curve correspond to the effective torsion equation of state as a function of $z$ expected for type II $f(T)$ theories with $\Omega_{\rm m0}=0.3$, $\Omega_{\rm r0}=8\times 10^{-5}$, and $n=3.5$.}
\end{figure}

\begin{figure}
\includegraphics[width=10cm]{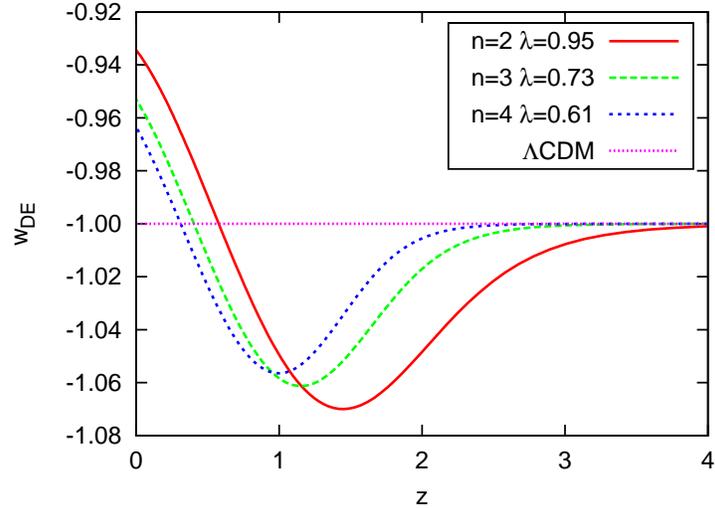}
\caption{The curves correspond to the effective equation of state as a function of $z$ as expected for Starobinsky's $f(R)$ model \cite{Motohashi}. \label{StarobinskyF}}
\end{figure}

\begin{figure}
\includegraphics[width=10cm]{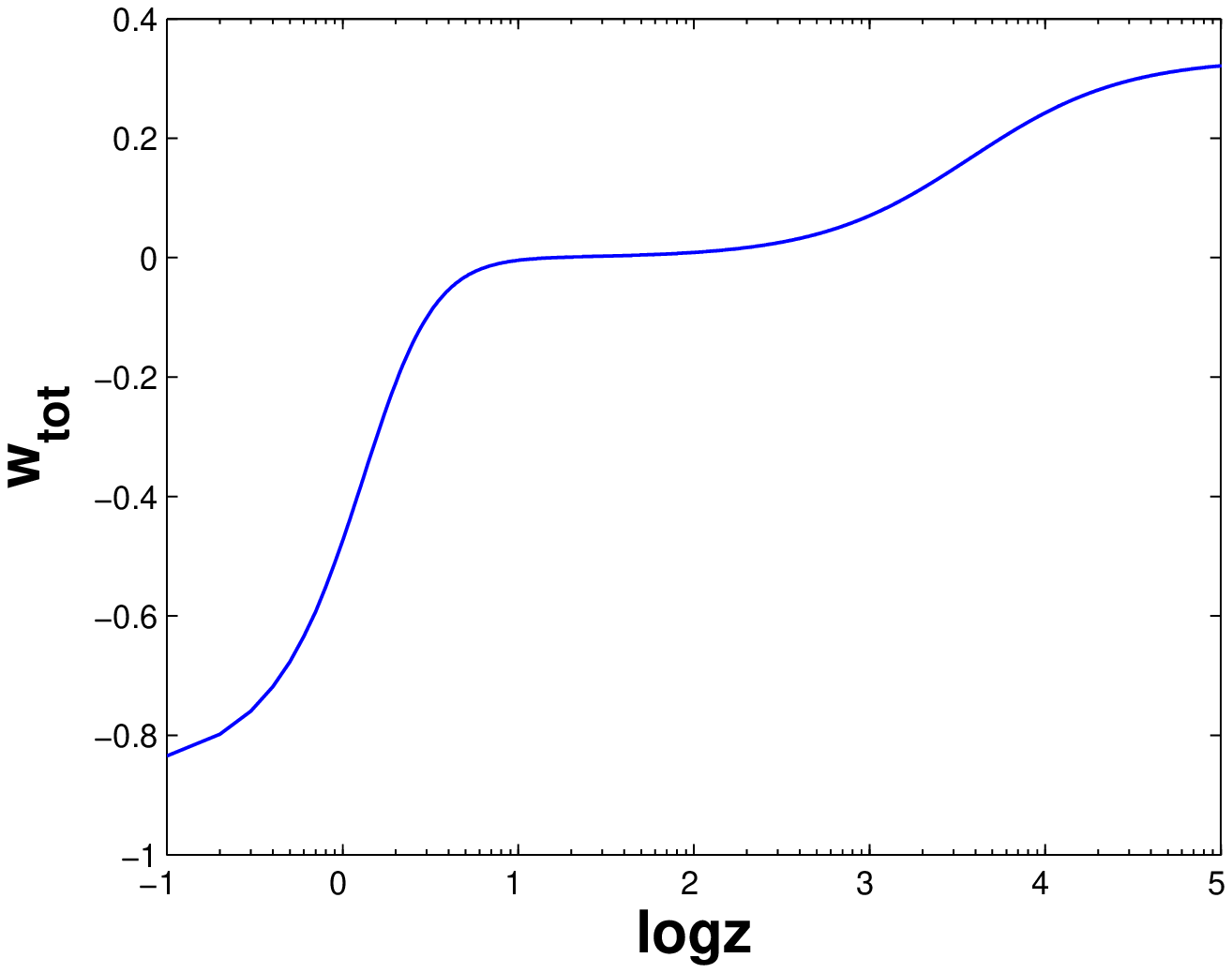}
\caption{The curve corresponds to the total equation of state as a function of $z$ expected for type II $f(T)$ theories with the same values of the parameters as in Fig. 4.
\label{Fig5}}
\end{figure}
\begin{figure}
\includegraphics[width=10cm]{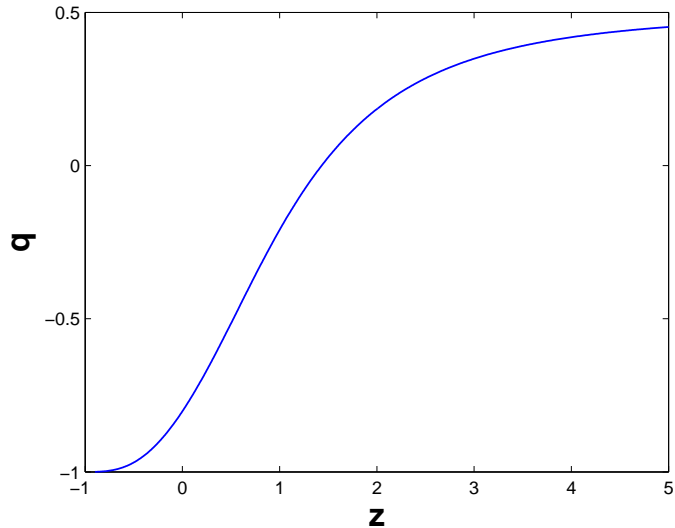}
\caption{The evolution of the deceleration parameter as a function of $z$ expected for type II $f(T)$ with the same values of the parameters as in Fig. 4. \label{Fig6}}
\end{figure}

\subsection{Type III: $f(T)=T+\alpha T_0\frac{(T^2/T^2_0)^n}{1+(T^2/T^2_0)^n}$}
Other teleparallelism alternatives for dark energy may be suggested along the same line. As an extension of the theory of teleparallelism, one may consider a model : $f(T)=T+\alpha T_0\frac{(T^2/T^2_0)^n}{1+(T^2/T^2_0)^n}$, in analogy to a $f(R)$ theory: $f(R)=R-\alpha R_0\frac{(R^2/R^2_0)^n}{1+(R^2/R^2_0)^n}$ \cite{Hu}, which satisfied both cosmological and local gravity constraints. The motivations for this type of $f(T)$ theory are: first, the behavior of the model should mimic $\Lambda$CDM at high redshift to be tested by the CMB; second, it should accelerate the expansion at low redshift with an expansion history that is close to $\Lambda$CDM, but without a true cosmological constant; third, to constraining small deviations from general relativity with cosmological tests, it should include the phenomenology of $\Lambda$CDM as a limiting case. This type of $f(T)$ theory can also behave like a cosmological constant, but because of its dynamic behavior, it is also free from coincidence problem.

The modified Friedmann equation, (\ref{f1}), for this type $f(T)$ theory reads
\begin{eqnarray}
\label{y3}
y+\frac{\alpha y^{2n}[4n-1-y^{2n}]}{(1+y^{2n})^2} =B.
\end{eqnarray}
For $z=0$, we have $\alpha=2(1-\Omega_{\rm m0}-\Omega_{\rm r0})/(1-2n)$ from Eq. (\ref{y3}). The values of $\alpha$ and $n$ have been related together through Eq. (\ref{y3}), so $n$ is the sole new free parameter, compared with general relativity. For $\alpha=0$ or $n=0$, the general relativity spatially flat Friedmann equation is retrieved. The case $T\rightarrow \infty$ recovers the general relativity dynamics with cosmological constant.

The effective energy density and pressure of torsion are given by
\begin{eqnarray}
\label{rho3}
\rho_T &=&\frac{6\alpha H^2_0 y^{2n} [1-4n+y^{2n}]}{2k^2 (1+y^{2n})^2}, \\
p_T &=& \frac{8n\alpha \dot{H}y^{2n} [4n-1-(4n+1)y^{2n}]}{2k^2 (1+y^{2n})^3}-\rho_T.
\end{eqnarray}
From Eq. (\ref{rho3}), the condition $\rho_T>0$ implies $\alpha>0$ and $1-4n+y^{2n}>0$, or $\alpha<0$ and $1-4n+y^{2n}<0$, and this imposes a constraint constraint on the parameter $n$: $n<0.25$. Otherwise, $\rho_T>0$ will be negative, which is unphysical. The parameter $n$ of the analogical $f(R)$ theory \cite{Hu}, however, is constrained as $n>0.9$ by the equivalence principle \cite{Capozziello2008}. This is the critical difference between this type $f(T)$ theory and the analogical $f(R)$ theory \cite{Hu}. Figure 8 shows the evolution of the effective torsion equation of state $w_{\rm eff}$ as a function of $z$ for this type $f(T)$ theory with $\Omega_{\rm m0}=0.3$, $\Omega_{\rm r0}=8\times 10^{-5}$, and $n=0.2$. $w_{\rm eff}$ decreases with redshift, reaches the max values at $z\sim 1$, and runs close to $-1$ in the future and in the past. For comparison, the effective equation of state $w_{\rm eff}$ of $f(R)=R-\alpha R_0\frac{(R^2/R^2_0)^n}{1+(R^2/R^2_0)^n}$ \cite{Hu} is also shown in Fig \ref{HuF}. According to their $w_{\rm eff}$, the difference between this type $f(T)$ theory and the analogical $f(R)$ theory is also obvious: $w_{\rm eff}$ in the $f(R)$ model \cite{Hu} can also cross the phantom divide, while that of the analogical $f(T)$ theory cannot. Figure 10 shows the evolution of the total equation of state. There can also be observed the latest three phases of the evolution of the universe: late acceleration ($w\rightarrow -1$), matter dominated ($w=0$), and radiation dominated ($w=1/3$). Figure 11 shows the evolution of the deceleration parameter. The transition from deceleration to acceleration occurs at $z\sim 0.9$.

\begin{figure}
\includegraphics[width=10cm]{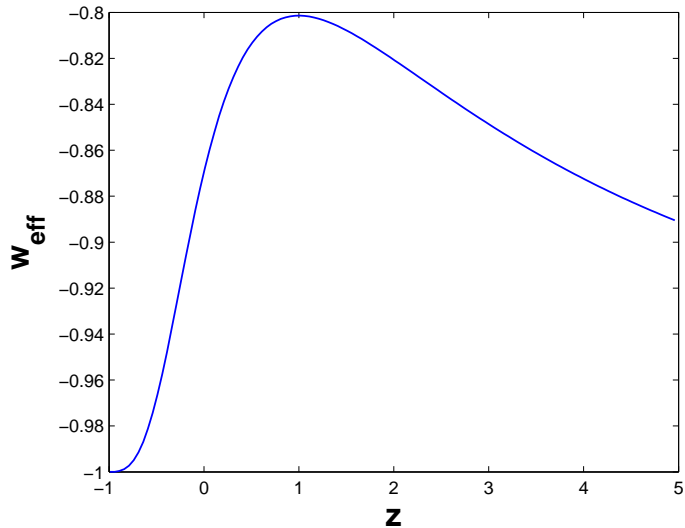}
\caption{The curve corresponds to the effective torsion equation of state as a function of $z$ expected for type III $f(T)$ theories with $\Omega_{\rm m0}=0.3$, $\Omega_{\rm r0}=8\times 10^{-5}$, and $n=0.2$. \label{Fig7}}
\end{figure}

\begin{figure}
\includegraphics[width=10cm]{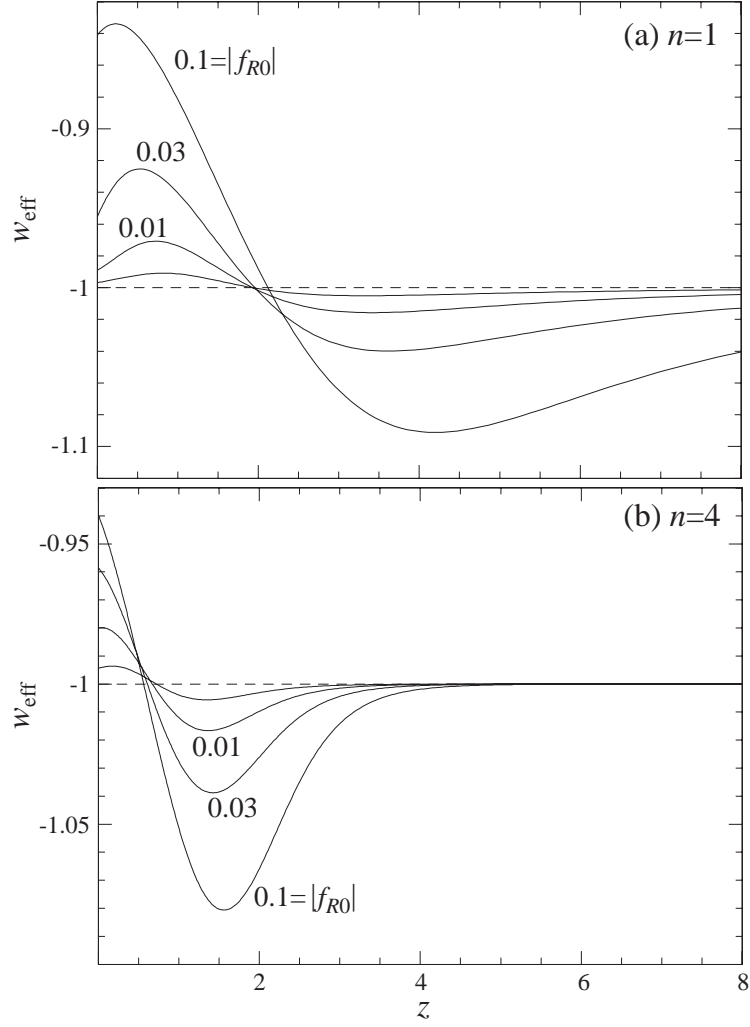}
\caption{The curves correspond to the effective equation of state as a function of $z$ expected for Hu and Sawicki's $f(R)$ model \cite{Hu}. \label{HuF}}
\end{figure}

\begin{figure}
\includegraphics[width=10cm]{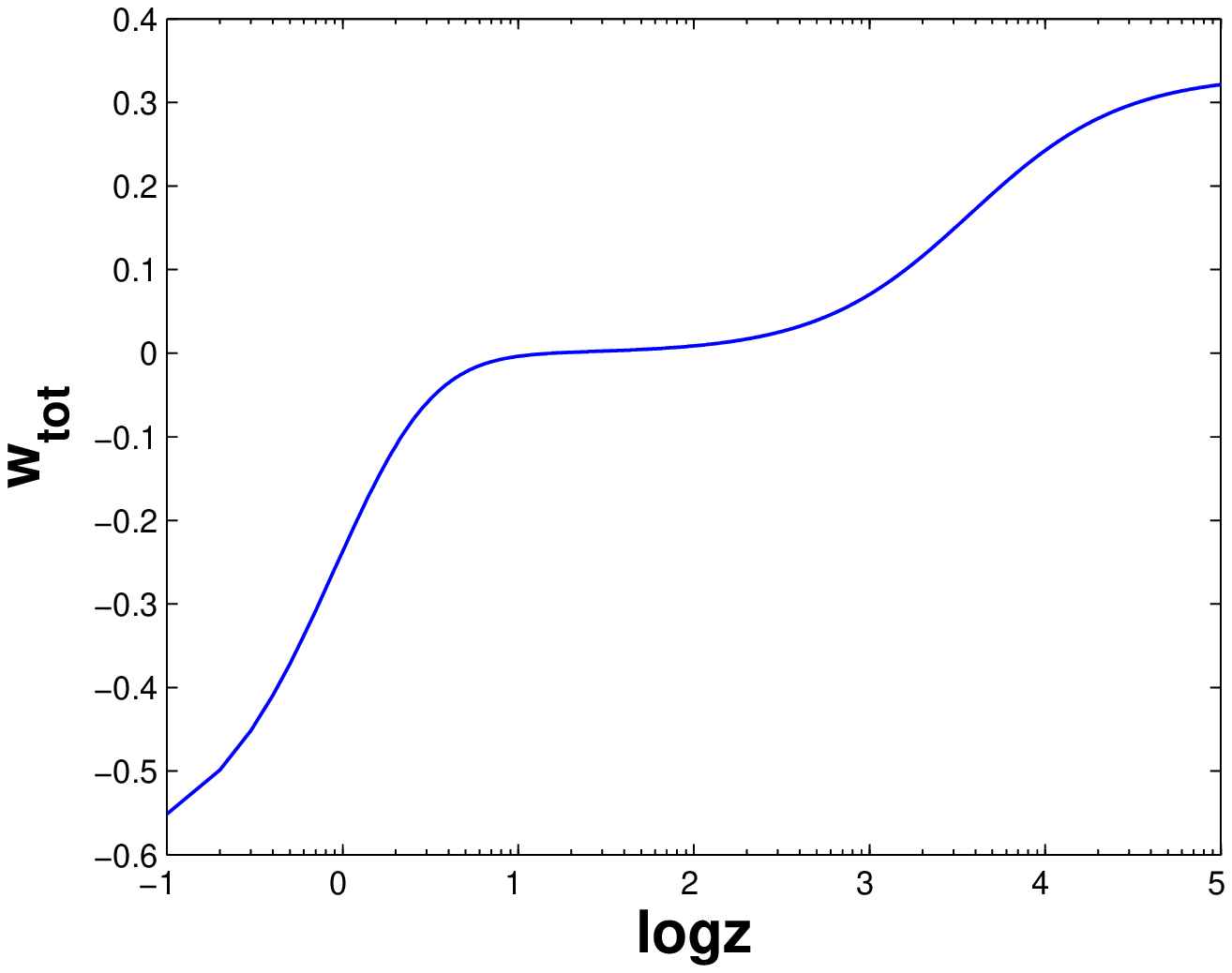}
\caption{The curve correspond to the total equation of state as a function of $z$ expected for type III $f(T)$ theories with the same values of the parameters as in Fig. 8. \label{Fig8}}
\end{figure}
\begin{figure}
\includegraphics[width=10cm]{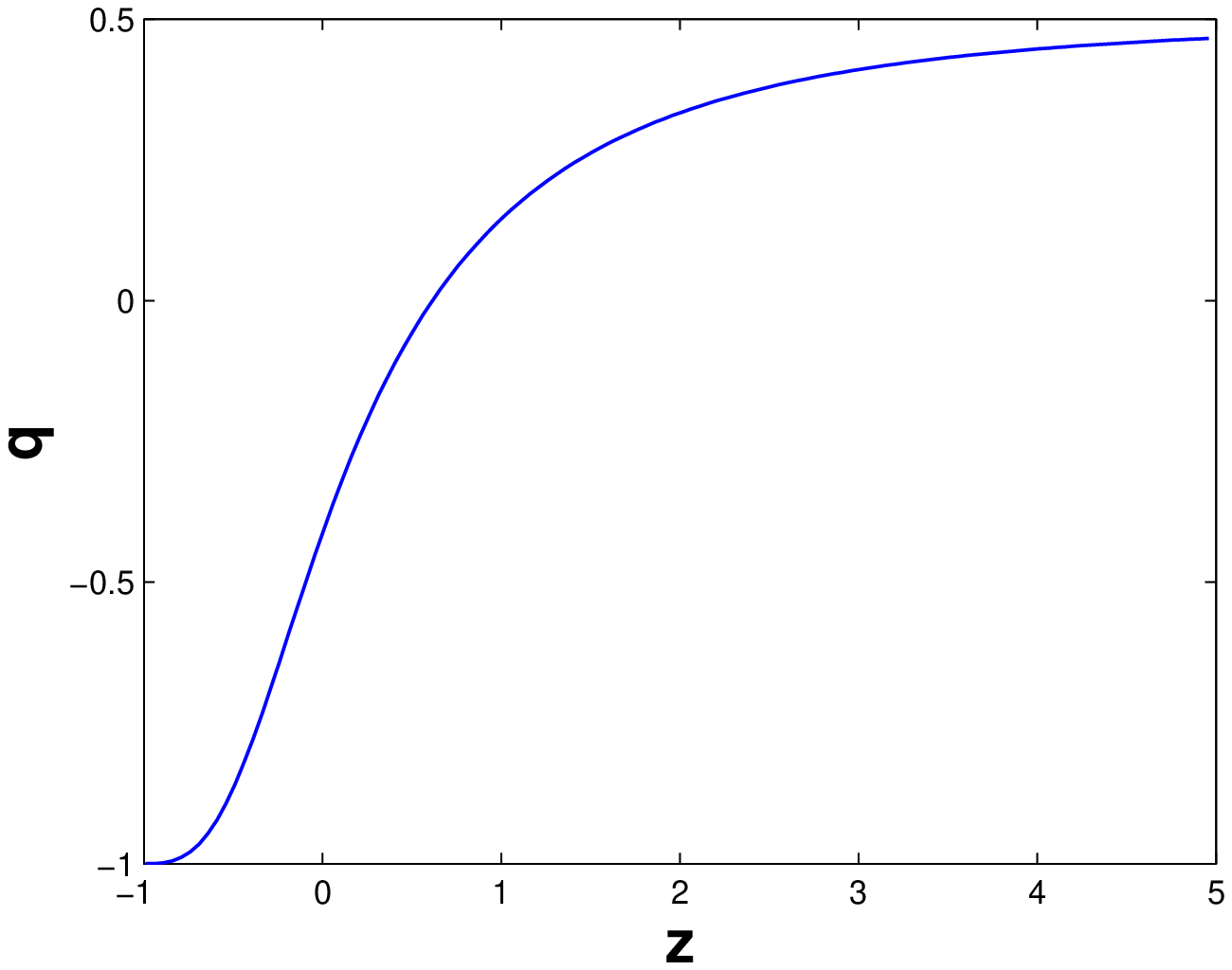}
\caption{The evolution of the deceleration parameter as a function of $z$ expected for type III $f(T)$ theories  with the same values of the parameters as in Fig. 8. \label{Fig9}}
\end{figure}
\section{Summary and conclusions}
The class of $f(T)$ theories, based on modifications of the teleparallel gravity where torsion is the geometric object describing gravity instead of curvature,
are in analogy to the $f(R)$ theories, but they have the advantage of second order field equations. We investigated three types of $f(T)$ theory here: I: $f(T)=T-\frac{\alpha}{(-T/6)^n}+\beta (-\frac{T}{6})^m$, II: $f(T)=T-\alpha T_0 \big[\big(1+\frac{T^2}{T^2_0} \big)^{-n}-1 \big]$, and III: $f(T)=T+\alpha T_0\frac{(T^2/T^2_0)^n}{1+(T^2/T^2_0)^n}$. Type I $f(T)$ theory can behave like phantom, which affords an alternative realization to obtain $w_{\rm eff}<-1$.
Type II and III $f(T)$ theories can behave like a cosmological constant, but because of their dynamic behavior, they are free from the coincidence problem. We discuss their potential physical implications, studied their cosmological behaviors by using numerical methods, plotted the curves of the effective torsion equation of state, the total equation of state, and the deceleration parameter. It was shown that all of them can give rise to cosmic acceleration with interesting features respectively, so they are worthy of further studies.

The connections between these three $f(T)$ theories and other $f(T)$ theory discussed in the literature are analyzed. The differences between these types of $f(T)$ theories and the analogical $f(R)$ theory are stressed. In future studies it will be necessary to constrain these $f(T)$ theories with observations (e. g., constraining by PPN parameters or gravitational waves as in \cite{Capozziello2009a,Capozziello2009b}), investigate their stabilities, discuss their singularities like the case in $f(R)$ theory \cite{Capozziello2009}, probe into their theoretical origins, etc. It is also essential to distinguish between these $f(T)$ theories by observational methods.

The analysis we performed indicates that these three types of $f(T)$ theories can be compatible with observations, but they just faces the problem from the
cosmological point of view, and their results can been taken into account only if these $f(T)$ theories successfully passe observational tests which is the subject of interest of other studies. Other topics concerning these models, such as large-scale structure and cosmological perturbations, are also crucial for assessing the viability of these theories as alternative explanations for the acceleration of the universe.

\begin{acknowledgments}
We thank Y. Liu for helpful discussions. This study is supported in part by the Hebei Provincial Natural Science Foundation of China under Grant No. A2011201147, Research Fund for Doctoral
Programs of Hebei University under Grant No. 2009-155, and by Open Research
Topics Fund of Key Laboratory of Particle Astrophysics, Institute of
High Energy Physics, Chinese Academy of Sciences, under Grant No.
0529410T41-200901.
\end{acknowledgments}

\bibliography{apssamp}

\end{document}